\documentclass[manuscript,screen]{acmart}
\AtBeginDocument{%
  }

\setcopyright{acmlicensed}
\copyrightyear{2025}
\acmYear{2025}
\acmDOI{XXXXXXX.XXXXXXX}
\acmJournal{CSUR}
\acmVolume{1}
\acmNumber{1}
\acmArticle{1}
\acmMonth{1}
\acmISBN{978-1-4503-XXXX-X/2018/06}

\begin{document}

\title{A Survey on the Use of Machine Learning within Computer Algebra}

\author{Uzma Shafiq}
\email{s224380786@deakin.edu.au}
\orcid{0000-0001-9146-2478}
\authornotemark[1]
\affiliation{%
  \institution{Deakin University}
  \city{Melbourne, Burwood}
  \country{Australia}
}
\affiliation{
  \institution{Coventry University}
  \city{Coventry}
  \country{United Kingdom}
}

\author{Matthew England}
\affiliation{%
  \institution{Coventry University}
  \city{Coventry}
  \country{United Kingdom}}
\email{Matthew.England@coventry.ac.uk}
\orcid{0000-0001-5729-3420}

\author{Nayyar Zaidi}
\affiliation{%
  \institution{Deakin University}
  \city{Melbourne, Burwood}
  \country{Australia}}
\email{Nayyar.Zaidi@deakin.edu.au}
\orcid{0000-0003-4024-2517}

\renewcommand{\shortauthors}{Shafiq et al.}

\begin{abstract}
This survey explores the use of Machine Learning (ML) in the field of Computer Algebra (CA), both to optimise existing CA algorithms, and to perform symbolic computation directly. Traditional symbolic methods, while mathematically rigorous, often are computationally expensive thus limiting their use in real-world applications. Recent advances have shown that data-driven techniques can address these limitations by guiding heuristic decisions, selecting optimal algorithms, predicting structural properties of algebraic objects, or even making direct symbolic computations. 

We take a systematic literature review approach and uncover work in CA applications including cylindrical algebraic decomposition, Gr\"{o}bner basis computation, symbolic integration, and many more.  The survey compares the different ML approaches that have been employed for these tasks, ranging from decision trees to transformers. Issues uncovered by the survey include the lack of benchmark datasets for CA, which hinders the comparison of methods and the generalizability of ML models.  The survey identifies the potential for explainable AI tools to help develop trust in decisions, and to drive forward CA research itself.
\end{abstract}



\maketitle

\section{Introduction}

Computer Algebra (CA) algorithms are deterministic procedures that are used to manipulate, simplify, and solve mathematical expressions. They have been used for carrying out many computational mathematics tasks such as symbolic integration and solving non-linear polynomial systems. Symbolic solutions are in contrast to numerical ones, which use approximations with floating point arithmetic.  Numerical methods are widely used and computationally efficient. CA on the other hand often pays for its accuracy with high computation costs, limiting its use to when absolute accuracy is crucial, or when we need the greater understanding of the underlying physical system afforded by the symbolic solution. 

Machine Learning (ML) is a subset of Artificial Intelligence (AI) that develops statistical models to draw inferences from data, allowing for systems to adapt without being provided specific instructions.  ML has found application throughout science and engineering over recent years, and underpins the prominent Generative AI tools so prominent today. The reader seeking background material on classical ML is referred to \cite{PRML_06}, on deep learning to \cite{DBLP_21}, and on reinforcement learning to \cite{RL_18}.  

This paper presents a systematic literature review covering the past 15 years on the topic of ML applied to CA.  The purpose of this survey is to support readers by highlighting key developments, current challenges, and future research directions.

\subsection{ML to generate CA}

There is a body of work in the literature seeking to use ML, specifically generative AI models, to generate the results of mathematical computation directly.  This work started in 2020 with the notable paper of Lample and Charton \cite{DLS_19} who trained transformers to integrate expressions and find analytical solutions to differential equations.  More recently, Kera et al. have trained a transformer model to produce Gr\"{o}bner bases of 0-dimensional ideals \cite{LCG_24}.  Even in this view CA is not replaced by ML: both of these papers make extensive use of existing CA tools to produce the datasets used to train the ML models, in the latter case defining a new algebraic problem for future study.  

This paradigm of direct computation is not easily integrated into a workflow in place of traditional CA.  Even exceptional models will not have 100\% accuracy, but the guaranteed exact computation is the ``\emph{unique selling point}'' of CA and most users would want to see this maintained.  Hence, most work in the literature to date seems to have fallen into a second paradigm of ML to guide CA, as we introduce next.

\subsection{ML to guide CA}

There are often choices that need to be made in CA algorithms which are not specified in the high-level algorithm description itself. For example, in Cylindrical Algebraic Decomposition (CAD), an algorithm to study non-linear polynomial systems, a variable ordering must be specified.  This variable ordering greatly affects the computational cost of the algorithm, but is often irrelevant for the analysis being undertaken.  Another example comes from Buchberger’s algorithm to compute a Gr\"{o}bner basis, a key tool to study a set of non-linear polynomial equations.  The algorithm processes a queue of data: the order in which the data is processed does not affect the correctness of the end result but can lead to computational paths of wildly different lengths.  Both of these algorithms are doubly exponential in the worst-case scenario, but could how we make these choices improve the average case complexity?  
Such choices in CA are commonly taken by either the user themselves, or a simple human-made heuristic destined by CA experts but often not subjected to scientific rigour or even scientific reporting. 

The majority of this review focuses on the use of ML for optimising CA algorithms by making choices that were previously based on human expert knowledge or human-made heuristics, since we find such studies more prevalent in the literature.  However, we will also cover those studies available that aim to replace CA algorithms entirely with ML-based approaches.  
We find examples for a range of CA algorithms in the literature.  We also see a wide variety of ML tools in use: from supervised learning models trained to make distinct choices for reinforcement learning agents that adapt strategies dynamically.

\subsection{Motivation}

The main motivation for this survey is that, while individual studies have demonstrated the efficacy of ML in specific CA applications, there is currently no unified resource that categorises and compares this work.  To the best of our knowledge, there is no similar survey in the literature.  The closest examples may be the position papers of the second author \cite{England2018}, \cite{England2022}; but these were short works, not systematic in production and containing just hand-picked examples.  As ML tools grow in complexity and capability, it becomes important to critically evaluate their applicability and impact within computational algebra, and to outline future opportunities and challenges. 

Furthermore, we have also introduced a taxonomy under different categories based on the application areas of CA, to understand the distinct approaches taken, although we will see some common issues emerge which we summarise at the end. 
This survey paper also highlights the limitations and identifies research gaps for future work. 

The paper continues in Section 2 where we outline the methodology employed for conducting the literature review. Section 3 presents the existing research on ML cylindrical algebraic decomposition, while Section 4 offers a detailed overview of ML for Gr\"{o}bner Bases. Section 5 explores additional applications of ML in other applications of CA. Finally, Section 6 summarises the key findings and discusses the limitations and research gaps in applying ML to CA.

\begin{figure}[ht]
\centering
\includegraphics[width=\linewidth]{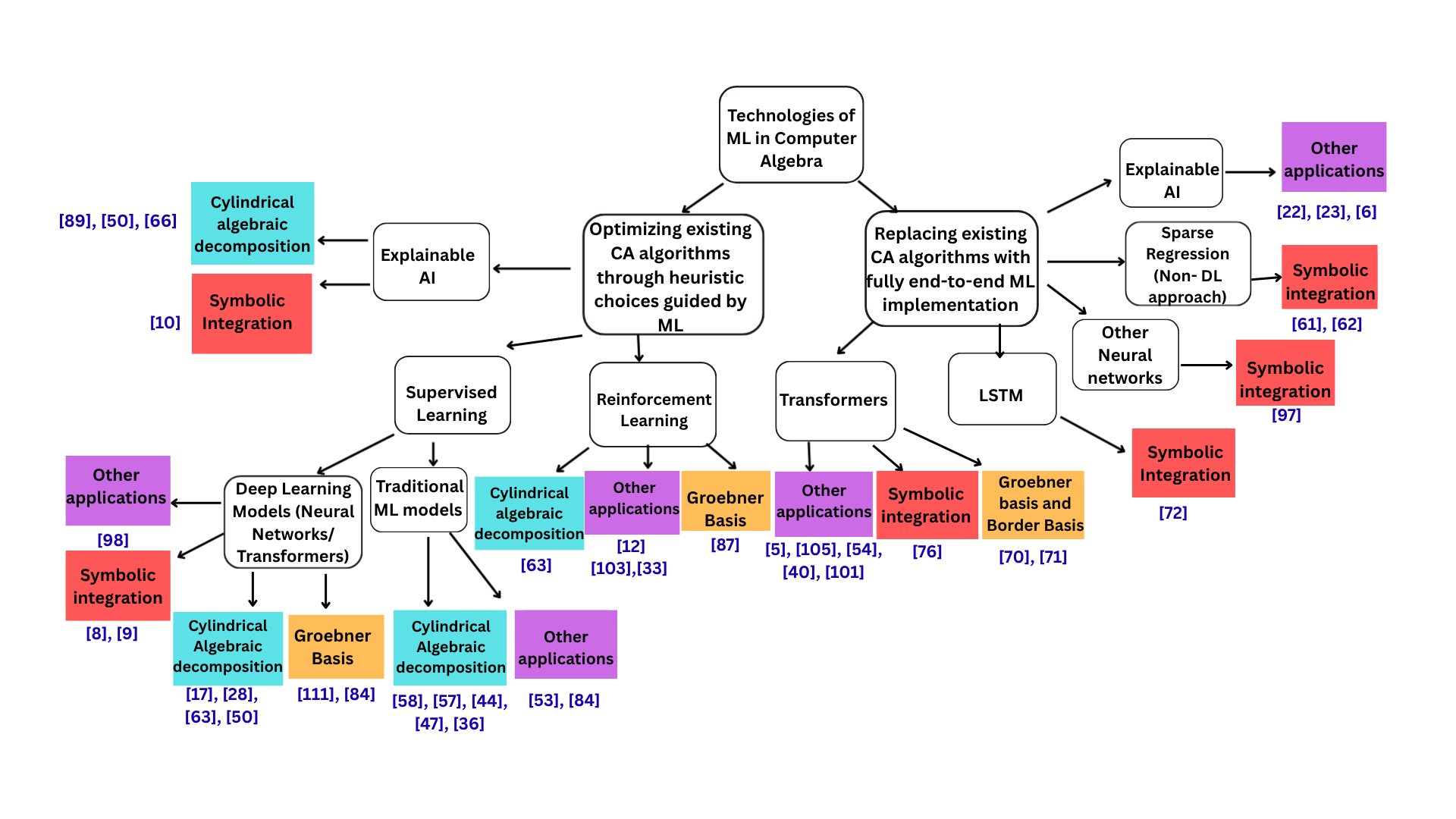}
  \caption{\label{fig:tax} Taxonomy for ML within CA employed in this paper.}
 \Description{A classification of literature based on ML tools and techniques.}

\end{figure}

\section{Literature Review Methodology}

We undertake a systematic  review of the literature over the past 15 years $(2010-2025)$. The motivation to select this timeline was that the earliest work applying ML to CA was 2014 \cite{AMLP_14}, so this range would capture that, and any others in years prior that we were unaware of.  
The review focuses on symbolic computation as the application domain only, and thus the use of CA to understand or improve ML technology is out of scope.

For the systematic search review keywords were divided into two categories. The first contained keywords to indicate the use of ML technology, and the second the application in CA (see Table \ref{tab:search_1}).  
We seek to capture applications throughout the field of symbolic computation, but we pay particular interest to non-linear polynomial systems where there is particular scope for efficiency improvements and so include many dedicated search terms here to ensure full coverage.

The databases that were searched were Scopus, IEEE Xplore, the ACM Digital Library and Science Direct. An additional separate search was carried out on arXiv to capture work in ML conferences and workshops which are not always formally published.  We also included a search on the online proceedings of MATH-AI workshop help over the last few years with NeurIPS and the AI for Math workshop at ICML. The initial results of the search are shown in Table \ref{tab:results_1}.

\begin{table}
  \caption{Keywords and databases used for searching relevant literature}
  \label{tab:search_1}
  \begin{tabular}{lp{10cm}}
    \toprule
    \textbf{Category} & \textbf{Keywords/Databases} \\
    \midrule
    Keywords for ML & Machine Learning, ML, Artificial Intelligence, AI, Reinforcement Learning, RL, Deep Learning, DL. \\
    
    Keywords for CA & Computer Algebra, Symbolic Computation, Non-linear Polynomials, Non-linear Systems, Non-linear Arithmetic, Gröbner basis, Groebner Basis, Grobner Basis, Buchberger's Algorithm. \\
    
    Databases Searched & Scopus, IEEE Xplore, ACM Digital Library, Science Direct. \\
    
    Additional Sources & arXiv, MATH AI Workshop Proceedings (NeurIPS). \\
    \bottomrule
  \end{tabular}
\end{table}

\begin{table}
 
  \caption{Number of papers by database and keyword combination}
  \label{tab:results_1}
  \begin{tabular}{p{0.2\textwidth} p{0.18\textwidth} p{0.22\textwidth} p{0.22\textwidth}}
    \toprule
    \textbf{ML Keywords} & \textbf{Symbolic Computation and Variants} & \textbf{Non-linear Polynomials and Variants} & \textbf{Gröbner Basis and Variants} \\
    \midrule

    \multicolumn{4}{l}{\textbf{Scopus}} \\
    AI variants & 347 & 222 & 89 \\
    ML variants & 104 & 192 & 20 \\
    RL variants & 9   & 80  & 1  \\
    DL variants & 27  & 76  & 3  \\
    \midrule

    \multicolumn{4}{l}{\textbf{IEEE Xplore}} \\
    AI variants & 65  & 37  & 1  \\
    ML variants & 63  & 78  & 0  \\
    RL variants & 3   & 3   & 0  \\
    DL variants & 8   & 34  & 0  \\
    \midrule

    \multicolumn{4}{l}{\textbf{ACM Digital Library}} \\
    AI variants & 43  & 8   & 2  \\
    ML variants & 30  & 6   & 3  \\
    RL variants & 6   & 5   & 3  \\
    DL variants & 3   & 2   & 0  \\
    \midrule

    \multicolumn{4}{l}{\textbf{Science Direct}} \\
    AI variants & 3   & 21  & 2  \\
    ML variants & 6   & 20  & 2  \\
    RL variants & 1   & 6   & 0  \\
    DL variants & 2   & 10  & 0  \\

    \bottomrule
  \end{tabular}
\end{table}

Of the total of 1,646 entries discovered, there were 585 duplicates. The studies were then filtered based on their titles and then their abstracts. Given the relatively limited number of studies identified in this area, a single-stage screening process was considered sufficient for the purposes of this survey. The results obtained were further filtered on the following criteria and a total of 76 studies were included: 

\begin{itemize}
    \item We searched for AI as this can be used instead of ML. However, we then screen out other forms of AI (e.g. theorem proving and SAT solvers may be considered AI but are not forms of ML). 

    \item Studies that concern the development of neuro-symbolic ML methods are not considered within scope: neuro-symbolic ML integrate symbolic reasoning into ML systems, the complementary direction to the ML for CA focus of this review.
\end{itemize}

The filtered studies can be classified in two ways: (i) based on the ML techniques that are employed, or (ii) by the CA problems that are addressed. While Figure \ref{fig:tax} presents a taxonomy organised by ML techniques, the survey follows the second approach. We found that organising the discussion by ML methods would result in excessive fragmentation and dense information, as many studies use overlapping or hybrid techniques. Structuring by CA applications provides a better understanding and clearer picture of how ML contributes to specific algorithmic improvements. The first taxonomy demonstrated in Figure \ref{fig:tax} is included to give readers an alternative perspective and to highlight the breadth of ML methods explored in this domain. However organising the studies according to this taxonomy in the survey would concentrate a lot of information in certain sections which could make it more difficult for readers to follow and appreciate the progress achieved across different CA applications. 

As we reviewed the remaining papers we found work focused in three main areas of CA: cylindrical algebraic decomposition, symbolic integration and Gr\"{o}bner basis computation.  We resent the work grouped in this way first, before a final section with the other applications identified in the literature.

\section{Cylindrical Algebraic Decomposition} \label{CAD}

Cylindrical algebraic decomposition (CAD) is a tool widely used for working with non-linear polynomial systems (both equations and inequalities) defined over the real numbers.  It was originally developed by Collins to solve the real quantifier elimination problem \cite{Collins1975}, where we are given a logical formula in real polynomial constraints that is quantified ($\forall, \exists$) and seek an equivalent formula without quantifiers.  E.g.
\[
\exists x, \, 
\bigl(a \neq 0 \;\wedge\; a x^2 + b x + c = 0\bigr)
\equiv 
a \neq 0 \;\wedge\; b^2 - 4ac \ge 0.
\]
A CAD is produced relative to input polynomial constraints so that each has a constant truth value within each of the decomposition cells, allowing an analysis of the system by analysing one sample point per cell. CAD decomposes the real space system into cylindrical cells:  a cell structure that makes it easy to test inclusion, to project and to invert cells --- facilities that we can combine to answer quantifier elimination problems.  For example, we solve the problem above by building a CAD over $(x,a,b,c)$ space, testing the truth of the formula in brackets at one sample point in each, project the true cells onto $(a,b,c)$ space, and combine their formulae to uncover the quantifier free formula on the right.   

CAD is a computationally intensive algorithm, doubly exponential in the number of variables in the worst-case scenario \cite{DH88}. This complexity limits the use of CAD for large problems. There are many factors that have been studied to improve the practical performance of CAD since its inception (the textbook \cite{CJ98} captured the first 20 years).  In the last decade a key development has been the integration of CAD within logic-based AI frameworks, specifically SMT-solvers for non-linear real arithmetic \cite{SMT_18}, and the systems that use them.  
This led to reformulations of the CAD theory, both to better fit, and taking inspiration from, the SMT context \cite{JdM12}, \cite{Brown2015}, \cite{ADEK21}, \cite{NASBDE24}. 

The CAD algorithm works by identifying additional polynomials over whose roots the truth of the input polynomials may change, and then building the decomposition according to these. Both the identification of the polynomials, and the building of the cells proceeds iteratively one dimension at a time.  There hence has to be a determined ordering of the variables for the algorithm to run.  To use the CAD for quantifier elimination we must order the variables as they are quantified, but this rarely determines the full ordering.  For example, in the example above we must order $x$ first but can have any ordering between $a,b, and c$.

\subsection{Heuristic decisions within CAD}

\begin{figure}[ht]
    \centering
    \includegraphics[width=0.40\linewidth]{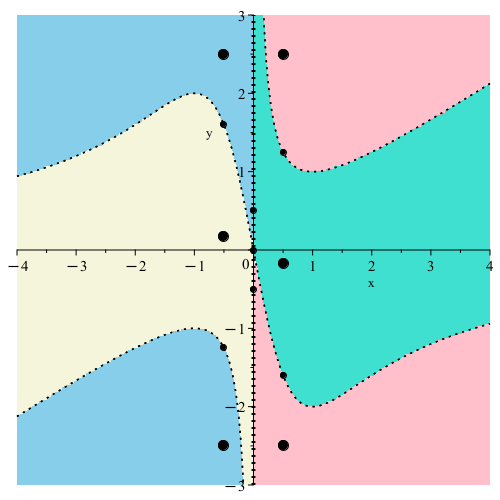}
    \includegraphics[width=0.555\linewidth]{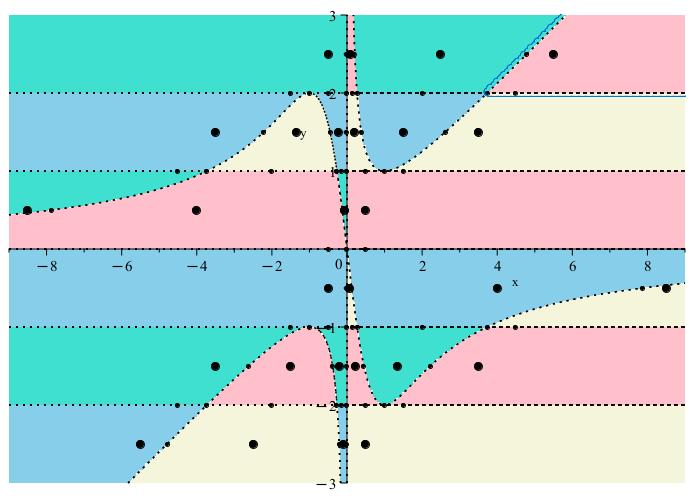}
    \caption{\label{fig:VO} CADs for \begin{math}\{x^3y + 4x^2 + xy, -x^2 + 2xy - 1\}\end{math}.  The left uses ordering $x \prec y$ and constructs 13 cells; the right uses $y \prec x$ and constructs 89 cells. The shaded regions show different 2D cells; and the different colour schemes the different cylinders. Sample points of each cell are marked with black discs:  larger ones for the 2D cells and smaller for the  lower dimensional cells.} 
    \Description{An illustrated example that uses different variable orderings for the same problem and shows its effect on decomposition.}

\end{figure}

As mentioned in the introduction, the variable ordering is one of the critical factors that affects the computational runtime of the CAD algorithm. There does not exist any exact method to ensure an optimal choice (short of trying all possibilities) and so in practice a heuristic choice is made if the user does not specify an ordering.  
Poor selection of the variable ordering can lead to a higher number of projection polynomials, explosive degree growth, and increased number of cells in the decomposition, and vastly different runtimes \cite{EPO_04}, \cite{CQE_07}. This is illustrated for a simple 2d example in Figure \ref{fig:VO} where one variable ordering results in 13 cells and the other 89, seven times more.

Due to the lack of an efficient and accurate tool, the choice of variable ordering is usually made by simple human-designed heuristics that use easily computed statistics of the input, such as \emph{Brown’s heuristic} \cite{usna} or \emph{gmods} \cite{NHC_22}.  However, none of these heuristics may be concluded strictly superior to the others $-$ all make mistakes for substantial subsets of problems \cite{AMLP_14}.  Due to the inadequacy of existing heuristics and the dramatic effect of the choice on the following computation, CAD variable ordering has become a key case study in ML for CA literature.

Another decision that has been studied in the literature is the preconditioning of the polynomial equalities present in the system using Gr\"{o}bner bases as first proposed in \cite{BH91}. A Gr\"{o}bner basis may be considered a simplification of the set of polynomials, and so it is perhaps not surprising that this usually benefits the subsequent CAD computation.  But that is not universally the case, and identifying the value in this preconditioning on a case-by-case basis can be beneficial.

Finally, some of the more advanced CAD algorithms act iteratively by polynomial, as well as by variable \cite{CM14b}, \cite{Brown2015}, \cite{NASBDE24}.  This introduces another heuristic choice, the order in which to process polynomials, which has received far less study than the variable ordering but can also have a significant effect \cite{EBCDMW14}.

\subsection{ML for CAD Literature}

A summary of all the relevant literature identified is given in Table \ref{tab:CAD}.  We discuss this thematically in the following sections.

\subsubsection{First systematic ML attempt}

The first attempt at using ML to select CAD variable ordering was made in 2014 by Huang et al.  \cite{AMLP_14}.  They used the NLSAT dataset from \cite{JdM12}: a set of satisfiability problems whose atoms were non-linear polynomial constraints.  The authors of \cite{AMLP_14} used only those problems involving three variables (7001 problems) and trained a simple ML model to choose which from three human-designed heuristics to follow for a given problem instance, seeking to minimise the cell counts in the resulting CADs.  They reported that the ML choice was superior to following any one heuristic.
Although the study opened a new paradigm for incorporating ML into CA, the proposed methodology was limited to the three variable case, and to the performance of the existing human-designed heuristics.

\subsubsection{Alternative formulations of the ML problem}

England and Florescu omitted the human-designed heuristics and had ML directly choose the variable ordering in \cite{CML_19}.  They again used the NLSAT dataset restricted to three variable problems, meaning the ML had a choice from 6 possible orderings.  This time ML models were trained to minimise CAD computation time: the ML tools outperformed the human designed heuristics.

Both \cite{AMLP_14} and \cite{CML_19} formulated the ML problem as classification (picking from a discrete set of options: human-designed heuristics in the former and variable orderings in the latter). One weakness of this approach is that the ML cannot take into consideration the wider ranking of orderings, i.e., even though the second-best ordering is not the correct classification it is preferable to the worst ordering.  The issue is particularly prevalent since the CAD run-times tended to gather in clusters (e.g. a cluster of orderings with similar quick run-times and another with similar slow run-times).  In this context, picking the right cluster is key.  
This problem was outlined in 2019 by Florescu and England \cite{ICV_20} who took a first step to address it by allowing the hyper-parameter selection in their workflow to be made with respect to actual runtime, rather than classification accuracy.  

In 2024 Del R{\i}o and England took the next logical step and reframed the ML problem as regression to predict the computation time with a given ordering \cite{LDP_24}.  This in turn solves the classification problem (choose the ordering we predict will be quickest).  In general ML regression is more difficult than classification, but here the key evaluation metric is not regression accuracy but the classification performance of the choices that regression led to.  The results demonstrated that regression could give better overall performance, but only when ML models best suited for this problem formation are used. 

\subsubsection{Fixing the number of variables}

A key weakness in the early work on this problem was the training of models for problems of a fixed number of variables, usually three (although in \cite{ICV_20} the authors increased to train a four variable model).  The number of different variable orderings is factorial of the number of variables, and so the decision becomes super-exponentially harder.  Although, since CAD itself is restricted by exponential complexity, there is a limit to how far we will be called upon to make such decisions.

This shortcoming was first considered in 2020 by Chen et al. \cite{CZC20}. The authors here proposed an iterative greedy method for generating candidate variable orderings and then an ML approach for selecting amongst these, allowing the ML to avoid the explosion in possible orderings. 

Also in 2020, \cite{BD20} confronted a similar issue of unbounded input size when considering the problem of choosing the order of polynomials for the adapted CAD algorithm \cite{Brown2015}.  The authors tackled this via forming an ordering of polynomials by repeated binary ML classification on pairs.  

\subsubsection{SMT-LIB dataset issues}

The most commonly used dataset in the work above comes from the \verb+NRA+ and \verb+QF_NRA+ categories of the SMT-LIB (\url{https://smt-lib.org/}).  This was seeded by the NLSAT dataset \cite{JdM12} and has been enlarged in the years since. These are problems primarily designed for solution by SMT-solvers and so lack problems with a wider range of quantifier structures that CAD can still study (for which there does not appear to exist any substantial dataset).  Nevertheless, the examples are mainly sourced from real-world applications, making them valuable for benchmarking purposes.  The library contains examples from fields as diverse as biology and economics, but the most prevalent examples are those created automatically by other software which calls an SMT-solver as a sub-tool, such as automated theorem provers and term-rewrite systems.  

However, all the work prior to 2023 described above, all overlooked biases present in this dataset: in particular for the three-variable subset that one particular ordering is far more often the optimal one than the other five.  This raises the question as to whether the ML models above are learning only this bias.  The issue was simultaneously reported in two 2023 papers \cite{AMD_23} \cite{DAM_23} who both proposed a process of data augmentation to address the imbalance by permuting variable names.  This creates new data instances, without any further cost in CA computation to label them, solving the imbalance problem and greatly increasing the dataset size.  The papers described how ML models  trained on a biased dataset do indeed perform poorly when testing on an unbiased one; but when trained on an unbiased augmented dataset then testing performance remains strong.  So the earlier models were learning the bias and removing that unfair advantage drops performance; but performance can then be boosted back by using the larger dataset gained through data augmentation.

The problems with the dataset go beyond this particular bias however.  It is described in \cite{LDP_24} that the dataset contains many problem instances which are very similar, although not identical, perhaps differing by one coefficient in a polynomial say.  This is caused for example by a theorem prover generating incremental SMT calls while developing a proof \cite{AP10}.  In many (but not all) cases such a small difference has no material impact on the resulting CAD calculation, which raises again the question of data leakage between training and testing.  The authors of \cite{LDP_24} address this by considering for each 3-variable problem problem the associated vector of six CAD cells counts (one for each variable ordering): where there are problems with identical such vectors all but one is discarded.  This reduced the dataset to a sixth of its original size, but because these extra problems did not add information, this did not affect the learning performance.

\subsubsection{Synthetic data}

The alternative to using a dataset of real-world problems is to create synthetic data.  This was first attempted for CAD in 2016, when Huang et al. addressed the question of whether to pre-condition CAD by replacing the equations present with their Gr\"{o}bner basis \cite{UMLT_16}. They generated 1200 random 3-variable problems, with moderate bounds on degree and coefficient size.  While they could train an ML model to address the question, there is no guarantee that real application problems will behave as the random data casting doubt over the generalisation ability.  
Synthetic datasets were also used in \cite{CZC20}, \cite{BD20}, and \cite{FE20-pipeline}, with the latter selecting the polynomial degrees, coefficients and number of terms from distributions whose mean and standard deviation matched the SMT-LIB dataset, to give greater chance of generalisability.  However, of those three only \cite{BD20} evaluated also on real-world data, which should be an essential step when using synthetic data.

We note that although random polynomial generation is cheap, synthetic data generation is still expensive if it needs to be labelled with CA for supervised learning.  As a result, none of the datasets discussed so far have been large enough to support deep learning methods.  This was addressed recently by Chen et al. in 2024 who used a high performance computing system to build a new  labelled synthetic dataset for CAD variable ordering \cite{DSVO_24} of more than 20,000 problems (which can be augmented to more than 300,000).  Further, these are all meaningful since for every example the runtime of the worst ordering is at least twice that of the best.  Aside from its use to study this problem, the dataset was also used in \cite{DSVO_24} to uncover bugs in CA systems! 

In addition to these works, two further datasets have recently been introduced. The first focuses on real quantifier elimination problems \cite{DRQ_26}, providing a benchmark for evaluating learning-based approaches in this setting. The second is an enhanced four-variable dataset designed for studying variable ordering selection in cylindrical algebraic decomposition (CAD) \cite{EFV_25}. Together, these datasets further expand the available resources for training and evaluating ML methods in CA.
 
\subsubsection{Feature engineered supervised learning}

A wide variety of ML tools were used in the above literature, however, due to the lack of a large-scale dataset until 2024, the work above used a feature-engineered supervised learning approach.

The original work of Huang et al. \cite{AMLP_14}, \cite{UMLT_16} used support vector machines (SVM), and did not test with any other ML models. SVMs are known to be sensitive to outliers and can become computationally expensive for larger datasets. 

The 2019 work of Florescu and England experimented also with K-nearest neighbour (KNN), Decision Tree (DT), and Multi-layer Perceptron (MLP) classifiers \cite{CML_19}, while the 2023 work of del R{\i}o and England included also XGBoost and the 2020 work of Chen et al. \cite{CZC20} and the 2024 work of John and Davenport \cite{EAM_24} included also artificial neural networks.  There is no common consensus on which of these models is best suited to the problem; it can vary with other factors.  In \cite{LDP_24} the KNN and SVM models performed better under regression than classification, while RF, MLP, and XGBoost performed worse for example.

These models all presume a human feature engineering approach.  The work of  Huang et al. \cite{AMLP_14}, \cite{UMLT_16} used a very limited number (11 for 3-variable problems) of hand crafted features, seeded from the statistics used in the Brown heuristic \cite{usna}. Such a small number of features likely leads to under-fitting.

To overcome this issue Florescu and England proposed an automated feature generation method from polynomials systems in 2019 \cite{AGNA_19}, again inspired by the type of operations used in Brown's heuristic but generating all possible statistics they can make (78 unique features for 3-variable problems; 7 times more than in previous studies like \cite{CML_19}, \cite{AMLP_14}).  The added features greatly improved classifier performance and this approach was taken up all subsequent work that used a feature-engineered architecture.  

\subsubsection{Reinforcement learning and graph neural networks}

One limitation in the reported literature so far is that supervised learning relies on datasets, expensive to label with CA and potentially prone to over-fitting. A study in 2024 by Jia et al. was the first to use Reinforcement Learning (RL) for the selection of variable ordering for CAD \cite{SVO_24}, where an agent was rewarded for its choices based on the number of CAD cells they led to.  Their approach allowed for a single ML model that could make decisions for systems of different numbers of variables, addressing another of the key shortcomings in the prior work. A Chinese language study in 2024 also used RL for CAD \cite{VOS_ch_24}.

Aside the moving to RL, \cite{SVO_24} was also notable as the first paper to make use of graph neural networks for the CAD variable ordering problem.  Graphs are formed from polynomial sets with nodes being variables and edges indicating that two variables appear in the same polynomial.  Such graphs have been previously shown to allow for the maintenance of polynomial sparsity during CAD \cite{LXZZ23}, and so this is a natural approach to try, although this approach was not reported as successful when used outside of the RL paradigm in \cite{EAM_24}.

The dataset used for training in \cite{SVO_24} was synthetic, randomly generated polynomial sets.  While the RL agents could significantly outperform other approaches on such data, their performance on the SMT-LIB was less dominant, indicating work still to be done to address generalisation.

\subsubsection{Deep Learning}
  
Recent research has explored the use of Transformer models to improve variable ordering selection in CAD \cite{BDB_26}. Since generating labelled data requires performing exhaustive CAD computations for every possible ordering, and there is limited labelled data available, this work proposed a pre-training and fine-tuning framework. The authors first pre-trained Transformer models on six specialized tasks, which force the Transformer to internalize deep polynomial structures and algebraic behaviours. These tasks included predicting exponent structures, algebraic feature vectors, and properties related to projection factors and resultants, all of which are relevant to CAD variable ordering, but cheap to derive data for. The pre-trained models were then fine-tuned on the expensively labelled CAD variable ordering datasets to specialize the models in predicting effective orderings.

The study showed that this approach outperformed traditional heuristic methods on public CAD benchmark datasets, including those from \cite{DSVO_24}.  The authors further introduced an enhanced four-variable dataset, \cite{EFV_25}, containing a significantly higher proportion of difficult instances with running times exceeding 100 seconds for training and evaluation. They also evaluated the approach on a separate SMT-based CAD benchmark dataset to show models trained on randomly generated polynomial can generalize to more realistic symbolic computation problems.

Although the pre-training and fine-tuning strategy improved CAD variable ordering prediction and achieved substantial runtime improvements over classical heuristics, the paper also observed that some pre-training tasks related to deeper symbolic operations, particularly those involving projection factors and resultants, were significantly more difficult for the Transformer models to learn accurately. This suggests that while the models can learn useful structural representations of polynomial systems, they do not yet fully capture the deeper symbolic reasoning required for more advanced algebraic computations.

\subsubsection{Explainable AI}
 
There have been numerous studies reporting the efficacy of ML tools for making the heuristic choice of variable ordering in CAD. However, the studies do not give any insight into why these ML tools are successfully making this choice. In 2024, Pickering et al. utilised the explainable AI (XAI) tool SHAP to gain insight into how the feature-engineered models were making their decisions \cite{XAI_24}.  SHAP analyses the impact of individual features, putting a number on their contribution to the overall decision \cite{LL17} and is perhaps the most widely used XAI tool.  In \cite{XAI_24} the authors aggregated these scores across the dataset and had models vote to produce an overall ranking of the automated features develop in \cite{AGNA_19}.  The top ranked features included some statistics already seen in the literature \cite{usna}, \cite{NHC_22} but also others that were new.

As a final experiment in \cite{XAI_24}, the authors sought to build a simple heuristic from three of the features, equivalent in complexity to the widely used Brown's heuristic \cite{usna}.  They studied all 120 possible triples and found the best, named T1, to consist of the top three ranked features by SHAP.  This suggest a piece of simple code (not involving any ML architecture) that could be easily  swapped for Brown's heuristic in any CA system.  The code is informed by XAI, but not data-dependent and of a complexity equivalent to a simple human-designed heuristic.  In \cite{DSVO_24}, a separate group tested T1 on an independent dataset and found strong performance, suggesting that XAI may be used to strip out complexity that leads to over-fitting, allowing for better generalisation between datasets.

Another XAI experiment in 2024 was conducted by Florescu and England \cite{CNNf_24}. Here the authors tried to train an inherently interpretable model (rather than explaining a complex one).  They also sought something of the complexity of Brown's heuristic, and so modelled this heuristic's behaviour with a small neural network and then used ML training to improve the networks performance.  The trained network could then be reinterpreted to code similar to Brown's heuristic.  This approach trained the network on random data but evaluated it on the SMT-LIB dataset, again demonstrating that a simpler, XAI informed, model can allow for better generalisation between datasets.

\subsubsection{Limitations}

This area of symbolic computation has seen the most study with ML methods to date, although it has been focussed on the variable ordering choice with the other heuristic decisions receiving very limited attention.  As noted in \cite{UML_19}, additional decisions also bring in the question of which order to make the decisions in: if taken together then the possibilities grow substantially.  There has been no study yet in the literature to propose a methodology for this.

There is also clearly a need for more work on the reinforcement learning approach, where there has been only two studies to date: since one of those studies also introduced the graph bases models we need an ablation study to understand how each innovation contributed to the performance gain.  We note that when considered the choice between regression and classification in \cite{LDP_24} the optimal choice of model and architecture changed and a deeper study is needed to understand what is optimal overall.

There has been substantial lessons learned over the use of datasets in this area, with the recent contribution in \cite{DSVO_24} opening the doorway to deep learning research.  But there still remains unanswered questions over how well random data represents real-world applications, and whether generalisation away from the training dataset is possible.  
The recent work with XAI methods suggests one route to offer this, but only two studies have been done to date and there is a need to both replicate the hypothesis and explore the wider range of XAI applications.

\section{Gr\"{o}bner Basis Computation} \label{gb}

Systems of non-linear polynomial equations appear in many areas of engineering, medicine and computer science. It is necessary to efficiently solve such systems for cryptography \cite{SGB_09}, robotics \cite{AR_10}, physics \cite{MD_07} and other fields. One of the main tools widely used to solve non-linear polynomial systems is the computation of a Gr\"{o}bner basis. A Gr\"{o}bner basis is a generating set of an ideal over a closed ring of multivariate polynomials \cite{HIN_75}. 

\begin{figure}[ht]
    \centering
    \includegraphics[width=0.48\linewidth]{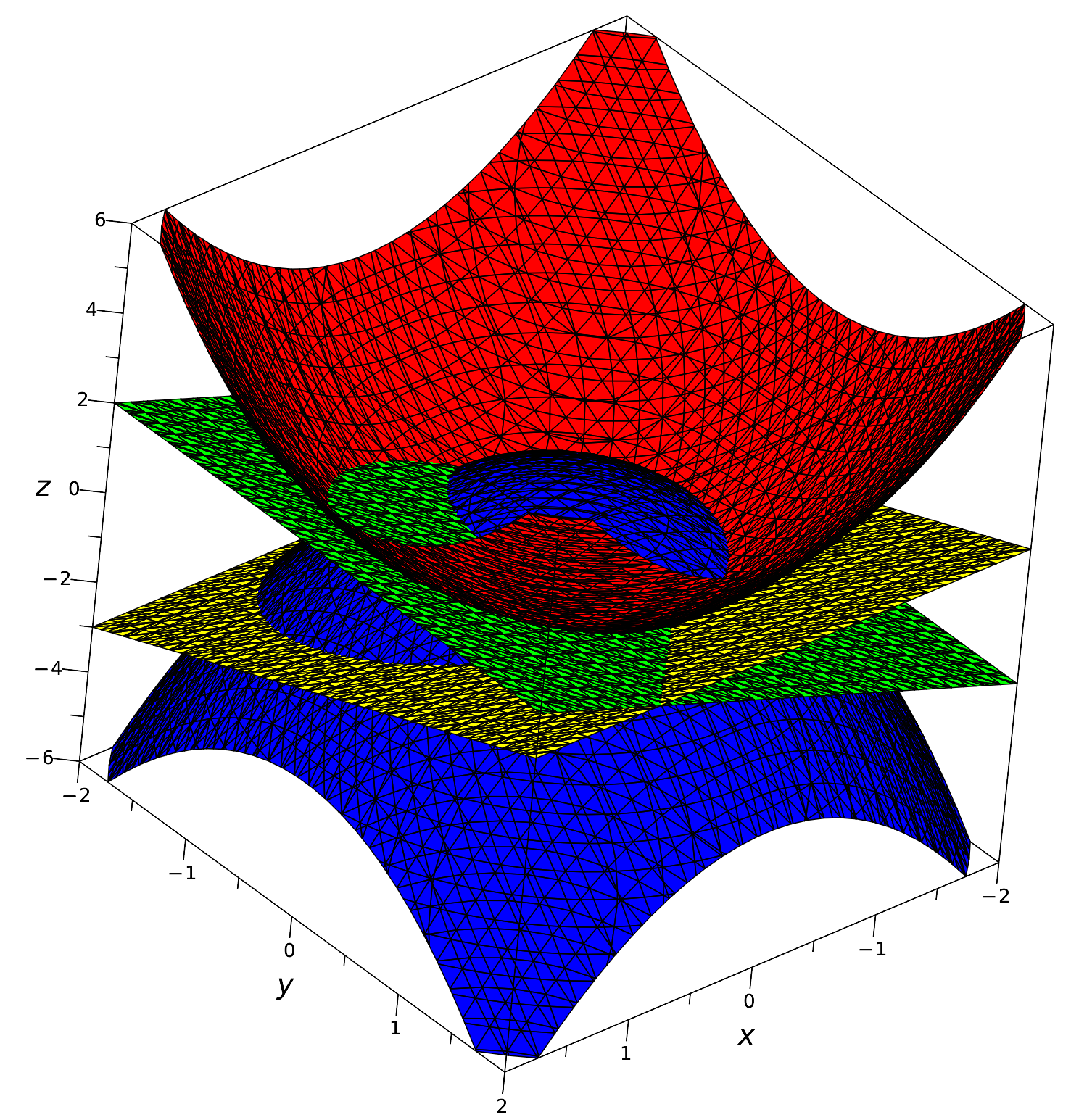}
    \includegraphics[width=0.48\linewidth]{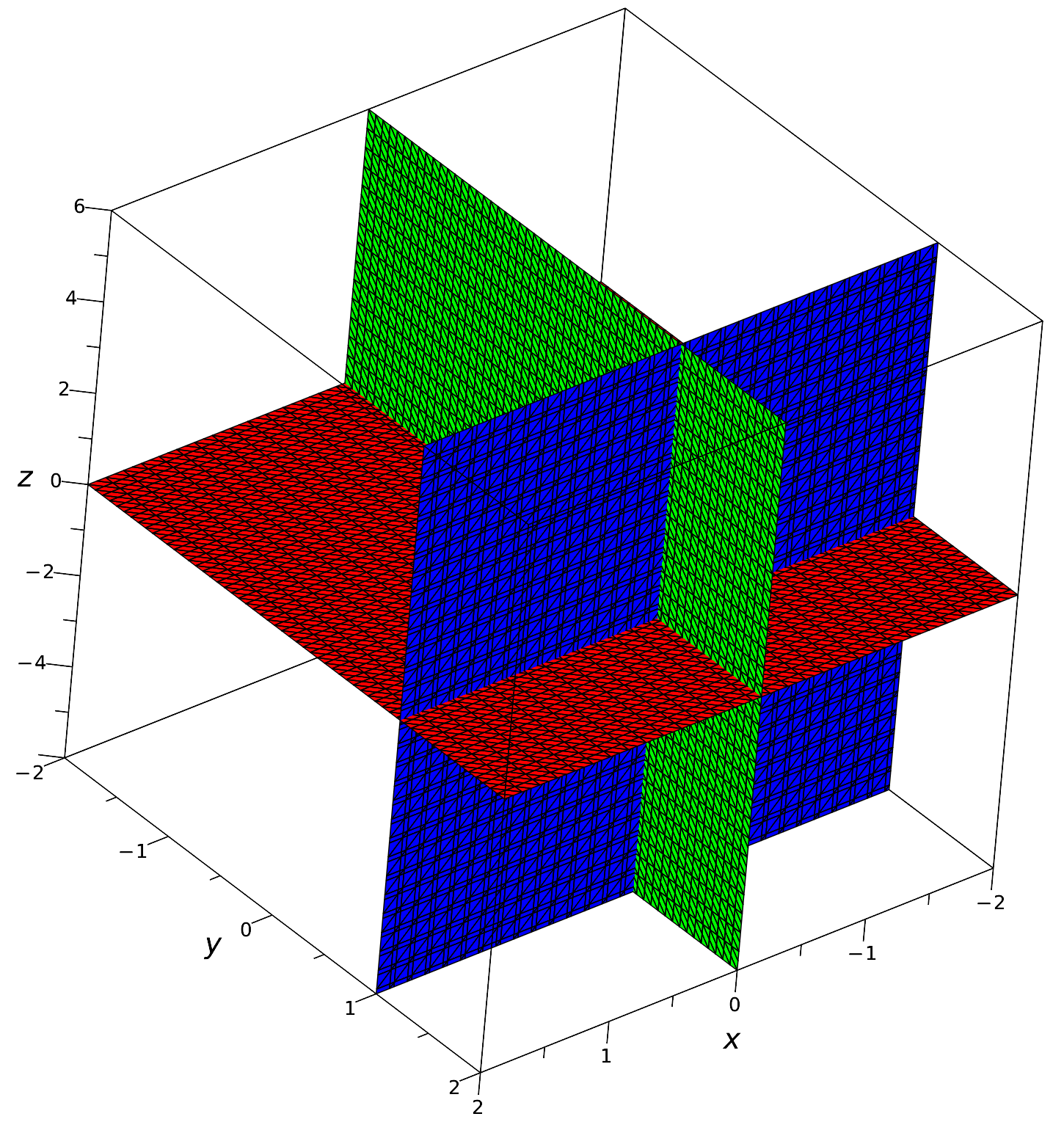}
    \caption{\label{fig:GB} An example in three dimensions (following  \cite{RoanesLozano2022}).  Let \begin{math} F = \{z-x^2-y^2+1, z+x^2+y^2-1, z-x, z-y+1\}\end{math} (as graphed on the left); then a Gr{o}bner basis for $F$ (under a lexicographical monomial ordering) is \begin{math}
    \{z, y-1, x\} \end{math} (as graphed on the right).  Note the curves on the left and right intersect at the same points.}
    \Description{An illustration of a Gr\"{o}bner Basis in 3D}
\end{figure}

Although ideals are infinite sets, they are generated by a finite ``generating set'' of polynomials. Interestingly, the same ideal can have different generating sets, which can reproduce the original generating polynomials through linear combinations. Some generating sets are more convenient than others: a Gr\"{o}bner Basis is a special generating set with advantageous properties. A Gr\"{o}bner Basis may be used to simplifies many computations and extract properties of an ideal.  It is particularly useful for solving systems of simultaneous equations represented by the ideal, making it a powerful tool in CA.  Consider the example in Figure \ref{fig:GB}:  the intersection of the four polynomials in the input is the same as the intersection of the three in the Gr\"{o}bner Basis, but the latter uses much simpler polynomials.

The most widely used algorithms for computing a Gr\"{o}bner basis are Buchberger’s algorithm (introduced in his 1965 PhD thesis \cite{Buchberger1965} and reproduced in English in \cite{Buchberger2006}), and Faug\'{e}re’s F4 and F5 algorithms introduced in \cite{Faugere1999} and \cite{Faugere2002} respectively. Both algorithms have doubly exponential complexity in the worst case scenarios \cite{MM82}, but the F5 algorithm is generally considered to be the fastest.  However, Buchberger's algorithm is simpler and easier to understand: the interpretability of that algorithm makes it a good choice for initial experiments on ML optimisation for improving  Gr\"{o}bner basis efficacy. 

The ML work in the literature to date has focused on optimising the internals of Buchberger's algorithm and so we briefly introduce this.  The algorithm begins with a generating set of polynomials for an ideal.  From this, a set $S$-pairs is initialised by all pairs from the input set.  Each time the main algorithm loop selects one of these, removing it from the set, with the loop terminating once the set is empty.  The loop body computes the corresponding $S$-polynomial by cancelling leading terms and reducing this with respect to the basis.  If the polynomial does not reduce to zero, then it is added to the basis and used to generate new $S$-pairs in the set by pairing it with the other basis members.  Pseudocode and an example for Buchberger's algorithm can be found in Appendix \ref{app:BA}.

\subsection{Heuristic decisions within Gr\"{o}bner basis computation}

When computing a Gr\"{o}bner basis there are choices that do not affect the correctness of the end result, but can impact how that result is presented and the resources used to obtain in.  First, like CAD, there is a choice of variable ordering needed, however, a Gr\"{o}bner basis is also computed relative to a monomial ordering.  It is widely thought that some monomial orders are more efficient to compute with than others: leading researchers to study algorithms to convert bases from one ordering to another \cite{CKM97}, \cite{Tran2000}. However, it has been observed that these general principles do not always apply to specific application domains \cite{SF19a}.  

A third decision that can affect Buchberger's algorithm is the order in which calculations are performed within, specifically the order in which $S$-pairs are reduced (as defined shortly).  Unlike the other two decisions this is not a single choice made one per algorithm call but a choice that needs to be made at the start of each iteration of the main algorithm loop.  The alternative GB algorithms also have heuristic choices in their calculation order worthy similar study:  the F4 algorithm uses linear algebra techniques to reduce large numbers of $S$-pairs at the same time,  but the batching still requires analogous selection strategies; and although the F5 algorithm selects pairs in order of signature it introduces a new sensitivity to the order in which the polynomials are presented (see e.g. \cite{CLO15}).

Work on these decisions to date (both ML and otherwise) has focussed on them individually, but of course, they can be taken in tandem: a classic work of Czapor described how while Gr\"{o}bner basis  with lex monomial orderings are hard in general, the problems can be mitigated through a good choice of $S$-polynomial selection \cite{Czapor1991}.

The choice of which $S$-pair to process next is critical for the computational complexity of the algorithm. Computation of $S$-pairs in a non-efficient sequence can lead to an increase in both the basis size, the number of loop iterations used, and the runtime.  There are well established selection strategies (heuristics) to make the choice, such as the sugar selection strategy \cite{GMNRT91}, but none of these is always dominant over the others (see e.g. section 4.3 of \cite{RLB_20}). The development of specialised strategies for particular applications remains a topic of current research \cite{KISYU23}.  
Given the prominence of this choice for the efficiency of the algorithm it has been viewed as a important candidate for ML optimisation.

\subsection{Machine learning for making heuristic choices for Gr\"{o}bner bases}
 ML has been implemented for selection of certain heuristics as well as end to end computation of Gr\"{o}bner basis. A summary of all the relevant literature identified is given in Table \ref{tab:GB} in Appendix \ref{app:ST}.  We discuss this thematically in the following subsections.
 
\subsubsection{Learning selection strategies for Buchberger's algorithm}

In 2020, Peifer et al. \cite{LSS_20} was the first one to apply ML tools to $S$-pair selection in Buchberger's algorithm.  They utilised Reinforcement Learning (RL) in which an agent (a neural network) makes choices and then receives a ``\emph{reward}'' to indicate how good the choice was $-$ the reward is used to edit the weights in the neural network to improve the next choice.  In this case the reward is the number of polynomial additions required to reduce the chosen $S$-pair.

They represent the current state (the current set of $S$-pairs) as a matrix, in which each row represents one pair by their exponent vectors (the powers of variables). Note that the coefficients are not considered.  They restrict their study to binomial ideals, meaning the width of the input matrix is fixes to a manageable size.   They also work in a modular field to avoid coefficient growth, and consider only the grevlex monomial ordering. Thus there is much work still to be done in generalising their approach to the full scope of Gr\"{o}bner basis calculations in applications. 

They both train and test with only random data.  They consider a variety of different distributions of random polynomials based on number of variables, number of generators, and degree. Their results suggest that agents trained on one distribution can perform also on other ``\emph{close}'' distributions.  

Later in 2021, Mojsilovic et al. used a similar setup to study the different parameters such as statistics of the input polynomial system and commutative algebra invariants derived from the ideal generators that can contribute towards the computational complexity of computing Gr\"{o}bner basis using Buchberger's algorithm \cite{LPM_22}. This time they seek to predict the number of polynomial additions a choice will require (i.e. the reward in \cite{LSS_20}).  The authors try first to use a multiple linear regression model built from a set of easy-to-compute statistics of the generator set, and later a recursive neural network both on these features and the polynomials themselves. They work only with binomial and toric ideals.

\subsubsection{Learning variable ordering}
 
Xu et al. in 2019 were concerned with the computational efficiency of Gr\"{o}bner basis for geometric vision problems \cite{OSI_19}. Here, solvers prepare fixed ``\emph{elimination templates}'' to calculate the Gr\"{o}bner basis online: they are concerned with selecting the variable and monomial ordering for a problem which leads to different templates.  They restrict the study to the case of choosing between different permutations of templates and train a deep learning model for the task.  Their model is trained on the basis of coefficients of polynomials to select the most efficient variable ordering, prioritising numerically stable results.  They train on synthetic data but evaluate on vision problems.

\subsubsection{Border basis optimisation}

A border basis is a generalisation of Gr\"{o}bner basis for the case of $0$-dimensional ideals (defining a finite number of points) which is  particularly useful in numerical polynomial algebra as it has better numerical stability compared to Gr\"{o}bner bases \cite{KK06}.  Kera et al. \cite{BBC_25} used transformers to select expansion directions in the algorithm to compute these; a similar choice to $S$-pair selection in \cite{LSS_20}, in that it needs to be made in each iteration of the main loop and the wrong choice can increase the runtime (but termination and output correctness is still guaranteed).  The transformer's choices led to speed-up factors of 3.5$\times$ compared to the base algorithm \cite{KK06}. 

\subsection{Learning to compute Gr\"{o}bner bases directly}

In 2024, Kera et al. utilised transformers to directly compute Gr\"{o}bner bases, showing the potential for transformers to directly address exponential complexity problems \cite{LCG_24}. Unlike \cite{LSS_20} this study is not restricted to binomials. but it does only consider $0$-dimensional radical ideals. 

The paper focuses on the problem of data generation: the power of transformers can only be utilised with large datasets whose production then becomes a CA problem in its own right.  They consider the random generation of a Gr\"{o}bner basis and then approaches to reverse engineer this into another generating set of the same ideal in order to produce pairs of input and output for a Gr\"{o}bner basis algorithm.  There remains work to do here since the data produced is quite biased in the relative lengths of the generator sets.

Another innovation in this paper is a new hybrid embedding of the inputs to overcome the problem of dimensionality: rather than have numerous number tokens (increasing the embedding matrix size) or splitting numbers into digits (increasing the attention heads), the coefficients go through a separate embedding network which encodes a continuity bias (two close numbers are close in the embedding).  The rest of the monomial (the variables and their powers) are stored as individual tokens, a different approach to the work in \cite{BBC_25} for border bases where each monomial was a single token.  Of course, all the data sampled still comes from fixed ranges.

Of course, the main limitation of this approach is that the results produced are not guaranteed to be correct, unlike those produced by symbolic computation methods like Buchberger's algorithm.  Note that although it is easy to check whether a given generator set is a Gr\"{o}bner Basis it is not easy to check whether it is the Gr\"{o}bner Basis for the original set.  This work requires new corresponding CA research to more efficiently check the validity of the transformer's answer.  Note that the group behind this paper have now released a Python library, CALT, designed to help non-experts in deep learning train models for symbolic computation tasks \cite{KAS25}.

Building on \cite{LCG_24} a new study \cite{HAT_25} utilised hierarchical attention transformers to compute the the Gr\"{o}bner Basis of multivariate polynomials. The tree-structured inductive bias better matches the hierarchical structure of polynomial systems and offers computational savings as compared to flat attention models. Furthermore by using  architecture combined with curriculum learning (training the model on datasets with progressively increasing problem sizes), the authors were able to solve problem systems that were much larger (13 variables)  compared to the ones reported in \cite{LCG_24} (5 variables). The study also provided evidence for generalizability to arbitrary tree depths. 

\subsection{Learning optimal input for Gr\"{o}bner Bases}

In 2020 a study demonstrated the use of ML tools to approximate the relations of highly complex polynomial systems with the aim of creating simpler problems for study with a Gr\"{o}bner basis \cite{FMN_20}. The context is modelling energy systems with many actors and finding Nash equilibria and this approach makes the proposed methods scalable to much more complex systems. However, the approximations might not be able to fully capture the complexity of the original systems.  Although the approach does not aim on optimising Gr\"{o}bner basis computation itself, it may be possible for the same techniques to be used to optimise the form of the input generating set in other contexts.

\subsection{Dataset challenges}

One aspect that becomes apparent in the studies above is the lack of a benchmark datasets for Gr\"{o}bner basis construction.  Without these, it is difficult to establish if any ML methodology is generalizable. The key studies above mostly both train and test on random data from the same generators, giving limited assurance against over-fitting. While the training and test sets are disjoint, when both are drawn from the same random generator they thus share the same underlying distribution and structural characteristics. Only a handful of papers partially offset this by experimenting with different distributions in the data generators.  We note the efforts of the Frisco, PoSSo and SymbolicData projects in the past \cite{Grabe2019}, the PHCPack database \cite{Verschelde1999}, and the more recent ODEbase \cite{LSR22}.  Nevertheless, a large-scale agreed community benchmark database for  Gr\"{o}bner basis, similar to the SMT-LIB would be a great asset.

\subsection{Explainable AI}

There has not yet been any application of XAI tools for Gr\"{o}bner bases in the literature.  Peifer et al. \cite{LSS_20} did offer interpretations of their agents’ strategy for the choice of $S$-pairs which seemed to have been obtained through manual analysis (see the thesis \cite{RLB_20} for details).  These revealed some preferences not previous considered in the literature:  preferences for pairs that had monomials, and for those whose $S$-polynomial had lower degree. Used alone, these preferences perform better than the previous strategies in the literature, suggesting the benefit of constructing simple heuristics from an XAI analysis as found earlier for CAD.

\cite{LPM_22} studied the contribution of complex algebraic features as well as simple statistics in their analysis to predict the number of polynomial additions performed.  Surprisingly, they found the simple statistics to outperform the more expensive algebraic invariants in terms of predictive power: this analysis is claimed to disprove folklore on the importance of certain invariants.  An XAI feature ranking approach might be able to add further insight on exactly which combinations of features are needed to make these predictions.

\section{Symbolic Integration} \label{SI} 

A summary of all the relevant literature identified is given in Table \ref{tab:symbolic_integration} in Appendix \ref{app:ST}.  We discuss this thematically in the following sections.

\subsection{Deep learning to compute symbolic integrals}

In 2020, Lample and Charton were the first to use transformers to directly undertake symbolic computations in \cite{DLS_19}, where they used transformers to perform symbolic indefinite integration and to solve ordinary differential equations.  Their results demonstrate that not only can transformers undertake such tasks with high accuracy, they can also tackle problems for which traditional CA systems cannot (at least within the time limit given: the experimental results report together failure due to timeout and production of the wrong answer, but the latter should come only from the transformer).  In many cases it is easy to check by differentiating but this is not always easy or possible and depends on the existing of high quality simplification tools.  

In \cite{DLS_19} the authors used sequence-to-sequence model, meaning both the input and output were viewed as a sequence of tokens (either number, variables, operators, or functions).  They constructed these sequences in prefix (Polish) notation for efficiency (avoiding brackets).  However, a subsequent work by Kubota et al. noted that prefix notation is not well suited to transformers since the operators and their subjects are not necessarily close to each other \cite{KTYF22}.  They experimented with other models and concluded that LSTMs, which learn the relative positions of tokens in the input, as superior for this task.

\subsection{Dataset issues}

The work in \cite{DLS_19} was replicated in a 2023 study \cite{SCD_23} but with narrower constraints for data generation and a larger dataset. 

The data is \cite{DLS_19} is generated from random expressions which are then either differentiated or integrated (using an existing CA system) to form training pairs.  Inadequacies in this dataset were the centrepiece of later critiques of \cite{DLS_19}.  Davis was the first to point out the pairs shared patterns in comparative length \cite{Davis2019}, with the dataset omitting pairs of roughly equal lengths.  This issue was addressed later in the works of Barket et al. \cite{GEI_23},  \cite{TLG_24}.  They proposed new data generation techniques for symbolic integration inspired by classical CA work for the Risch \cite{Risch1969} and Risch-Norman algorithms \cite{NM77} respectively.

Another concern with the dataset in \cite{DLS_19} is that the training data comes from the same generation methods of the testing data.  The authors did report how models trained on data from one generator performed when tested on another and identified generalisation issues.  In the extended version of the paper \cite{PBUK19} the author's analysis of the dataset in \cite{DLS_19} reveals that many of their test expressions (up to 74\%) are very similar, meaning they differ only a constant or sign, to the training data.  This particularly calls into question the appropriateness of the testing methodology.   

Concerns may be partially addressed by the additional data generators in \cite{GEI_23}, \cite{TLG_24} and by merging similar examples, but ideally we want an independent validation dataset.  Once again, there are not agreed community benchmarks in use for symbolic integration.  However, in this case, there does exist a large ($>72k$) set of problems produced for the RUBI system \cite{RUBI}.  The work in \cite{SIA_24} also made use of a large-scale validation set used by Maplesoft as unit tests for the Maple CA system.

\subsection{Learning for guiding symbolic integration algorithms}

In 2024, Barket et al. researched the problem of choosing a suitable method for indefinite integration of a expression in the CA system Maple \cite{SIA_24}.  There were 12 methods available and the current Maple user-level command tries them in the same fixed order until one succeeds, in some cases with simple ``\emph{guard}'' code to check applicability.  This wastes runtime, but can also lead to unnecessarily complicated output: different algorithms can produce different (but mathematically equivalent) integrands.  Barket et al. tackled this as a ML classification problem to select the sub-algorithm that produces the output with smallest size under Maple's data storage format for expressions.  They experimented with long short-term memory neural networks (LSTMs) for the task.  They found performance improved when using LSTMs for tree structures rather than just simple sequences, perhaps unsurprising given mathematical expressions naturally form trees, but a technique not used before in the literature.  They trained and tested on the data generators in \cite{DLS_19}, \cite{GEI_23}, and \cite{TLG_24}.  They also evaluated on an independent dataset which showed issues of generalisation still to be addressed.

A later paper by Barket et al. in 2024 looked at the guard code mentioned earlier:  these are simple pieces of code that can identify the applicability of an integration method for an integrand that exist for 4 of the 12 methods in Maple \cite{TPA_24}.  They trained transformers to make binary predictions on a methods applicability: these proved successful in generating guards for methods without one, and more accurate that the existing guards for methods which had them.  

In 2024, Sharma et al. considered a similar problem of ML guided integration, however, in their case they sought to apply rules to perform integration step-by-step as a student would by hand \cite{sird_23}.  They created a dataset containing 2 million pairs of mathematical functions, and integration rules that can be applied to them.  They use this to train a neural network to guide the choice of integration steps.  There is no risk from incorrect results: if the network suggests an inapplicable integration rule then this is identified.  They report that this leads to $6\times$ fewer steps when integrated into SymPy's \verb+integral_steps+ function.  The data was generated by runs of the original function, but the model can now guide the function to tackle problems it could not before, suggesting some clear generalisation has occurred.

\subsection{Frontier LLMs applied to symbolic integration}

Increasingly we see frontier models, i.e. LLM based AI systems such as ChatGPT, Claude, DeepSeek, evaluated on mathematics problems.  The MATH dataset included symbolic integration as a category \cite{MMP_21}, and became a focus for LLM research, see e.g., \cite{NNS_22}, particularly for so-called reasoning models following a chain-of-thought process \cite{WWSBIXCLZ22}. The MATH dataset contains step-by-step solutions for its problems, and so this is more like the task of Sharma et al. \cite{sird_23} than the direct symbolic computation of Lample and Charton \cite{DLS_19}.  However, unlike \cite{sird_23} the LLMs directly produce the steps without validation that they are correct.

\subsection{New symbolic-integration method utilising spare regression}

Irvanian et al., added symbolic integration functionality to the SciML package for Julia via a hybrid algorithm that included some learning techniques \cite{SNI_22}, \cite{ISG_24}. They were  inspired by the Risch-Norman algorithm \cite{NM77} to generate an ansatz with multiple candidate terms for the integral (which they do with a combination of table lookup rules and algebraic manipulations).  They then filter the candidate terms and apply sparse regression to find the coefficients. The approach has the benefit of finding integrals in a form similar to the provided integrand (what users likely expect).  

\subsection{Explainable AI}

There has been very little use of Explainable AI in this area.  Note that although \cite{sird_23} produces a method of integration that is interpretable to the user, unlike say the work of \cite{DLS_19}, the actual ML methods used are not explainable, i.e. we do not know why the neural networks makes the recommendation of integration rule that it does.  

The only paper to use an XAI method was \cite{TPA_24} who used layered integrated gradients (LIGs) to analyse the transformer guards for integration methods.  LIGs is a model specific post-hoc explainability technique providing local explanations \cite{AAD_17} by highlighting the importance of individual input features (tokens in this case).  For the most part this did not explain the transformers in \cite{TPA_24} (where of course it is likely combinations of tokens that need to be analysed) but this did reveal one interesting finding:  a particular function token which heavily contributed to predictions that a method should not be used.  Upon inspection it was discovered that this function is indeed not in scope of this method: the XAI essentially revealed a shortcoming in the existing guard code which could be easily patched in the future.

\section{Other applications of ML in CA} \label{OA}

A summary of the remaining relevant literature that we identified is given in Table \ref{tab:combined-symbolic}.  This includes some work similar to that we have seen in the previous sections but for different applications, which we summarise in the first few sections below.  We also finish by noting some of work that does not directly perform symbolic computation but nevertheless may have useful lessons to learn from.

\subsection{Using ML to guide symbolic algorithms}

In 2023 Berczi et al. \cite{ARS_23} considered how ML can help with the resolution of singular points, where we replace such points in the solution set of a system of polynomial equations with smooth ones whilst keeping the rest of the solution set unchanged.  Resolutions are not unique: the usual way to describe them involves repeatedly performing a fundamental operation known as ``\emph{blowing-up}'', with the complexity of the resolution highly dependent on certain choices.  They translated the problem into a two-player game such that a winning strategy for the first play provides a solution to the resolution problem.  They use reinforcement learning to train an agent to win the game and thus find the optimal resolutions.  The agents trained outperform state-of-the-art heuristics.  The reward is calculated as the total number of polynomial additions performed, similar to the work of Peifer et al. for optimising Buchberger's algorithm \cite{LSS_20}. 

ML has recently been used to guide coordinate transformations for polynomial ideals: \cite{MLPS_24} considers the problem of putting ideals into quasi-stable position, which is a special coordinate system that retains many desirable properties of the generic initial ideal while being deterministically computable and verifiable. The algorithm which exists for this is under-specified: it needs to make a choice of \emph{elementary move} to apply with any choice acceptable in terms of correctness and termination but giving different levels of efficiency. Traditional methods to make this decision rely either on random coordinate changes, which destroy sparsity properties, or simple human-designed metrics, which do not always yield optimal results. The authors explore a data-driven approach based on supervised ML with feature vectors of simple statistics about the ideal.  Their experimental results demonstrate that trained models frequently identify nearly optimal transformations, preserving structure and improving computational efficiency.  However, they both test and train with random ideals which, as they acknowledge, leaves open questions of generalisability.

\subsection{ML algorithm selection for resultants}

We saw earlier studies on integration algorithm selection for particular problem instances \cite{SIA_24}, \cite{TPA_24}.   A similar study was carried out earlier for selecting among different algorithms to compute the resultant of two multivariate polynomials \cite{AASC_16}. Simpson et al. used a feature engineering supervised learning approach for choosing from the four possible choices in Maple and Mathematica.  They experimented with a variety of ML methods, finding neural networks performed the best: they reported 68\% runtime improvement in Maple and 49\% improvement in Mathematica.   This study is a proof how ML can be used to make traditional CA systems more efficient.  Furthermore, investigating these black-box models using XAI might result in new insights for making this choice and corresponding new algorithm development.

\subsection{ML to improve CA for physics}

There is a long history of dedicated CA systems for physics calculations, often tailored to performing a small number of calculations many times at high speed.  These systems are highly optimised and so early adopters in the use of ML methods for CA optimisation.  A key optimisation that has been studied in \cite{KPVv13}, \cite{CSA_14}, \cite{KUV15} is to construct polynomials in a form optimised for speedy numerical evaluation, by optimising the order of variables used in Horner factorisation.  The authors use variants of Monte Carlo Tree Search to learn this, including in combination with simulated annealing, for the specialised CA system FORM.  

More recently, Alnuqaydan et al. studied how ML can replace the symbolic computation in the computation of squared amplitudes, a key task in high energy physics calculations \cite{SYMBA_23}. The authors use a sequence-to-sequence transformer model to replace this most expensive part of the symbolic computation: they report that ML is able to predict this correctly $97-99$\% of the time, at a speed that is up to two orders of magnitude faster than the symbolic computation.  It is not clear what the dangers are to the wider physics experiments from the small number of error cases of the ML.

\subsection{Linear polynomial equation solving}

We are mostly concerned with non-linear polynomial systems, but note that linear programming is also being addressed with ML tools.  Of particular interest is the paper of Dabelow and Ueda in 2024 which proposed a methodology to solve linear polynomial equations using deep neural networks and a reinforcement learning paradigm \cite{SER_25}. They utilise a symbolic stack calculator that applies different transformations to solve the equations.  The RL agent can suggest actions, such as applying mathematical operations.  The framework ensures that poor choices from the agent do not lead to incorrect mathematics (although they may not lead to the solution of the problem).  This framework does not make use of existing CA tools: if it had then it is possible these would have allowed it to improve performance, but on the other hand that would have limited it to strategies already designed by humans.  They emphasise that the solutions provided are comprehensible to humans and step-by-step (although like \cite{sird_23} the decisions of the actual agent are not themselves explained).  This work sits in contrast to approaches like \cite{VO24} which seek to train transformers to solve systems of linear equations directly.  

\subsection{Learning from the MATH-AI community}

As mentioned earlier for integration, we increasingly see frontier LLM models applied to mathematics \cite{MMP_21}, \cite{NNS_22}, including some tasks traditionally thought of as symbolic computation. A survey in 2022 on performance of LLMs on basic arithmetic concludes that state-of-the-art LLMs fell short when probed with relatively simple tasks \cite{CNND_24}. However, another survey in 2025 \cite{SLLM_25} demonstrated substantial improvements in the mathematical reasoning capabilities of LLMs through approaches such as Chain-of-Thought reasoning, reinforcement learning, test-time scaling, and knowledge-augmented reasoning. Despite these developments, current state-of-the-art models still face important challenges related to reasoning efficiency, exploration diversity, optimality of reasoning paths, and robust generalization to more diverse domains \cite{SLLM_25}.

Success in more advanced mathematics has come from more specialised architectures, such as AlphaGeometry which tackled Olympiad problems through a combination of reinforcement learning a deep neural network and a symbolic deduction engine \cite{TWLHL24}; or its sister project AlphaProof which generates formally verifiable mathematical proofs in Lean, again using reinforcement learning with feedback from a check using Lean \cite{AlphaProof}.  It has already been proposed that a similar joint approach is needed for symbolic computation and LLMs \cite{Wolfram2023}.  CA systems like Mathematica now integrate LLMs throughout their product, but there is not yet published research on the right way to optimise this.

There is a growing ``\emph{Math-AI}'' community looking at specialised AI models for mathematical tasks.   The focus here does not tend to be direct symbolic computation but the methodological developments may well have use for that.  For example, the work using tree-embeddings for symbolic integration method selection described earlier \cite{SIA_24} followed work using the tree-embeddings for mathematical formula search in \cite{WLB21} and for recognising the equivalence, derivative, or variable substitution between pairs of polynomials in \cite{LMR_21}, which may be useful for symbolic simplification.  The latter stressed the importance of the synthetic data generation for their work and credited this with much of the improved performance.  A related architectural development that used tree structures was reported in \cite{HAT_25}, where Gr\"{o}bner bases were computed using Hierarchical Attention Transformers, where tree structure is built into the attention mechanism itself rather than just the input embedding, yielding substantial efficiency gains and scaling to larger polynomial systems.

\subsection{Explainable AI}

None of the above work made use of explainability for the actual ML.  There is very limited work on XAI for LLMs, although we note the recent report from Anthropic:  in particular, this seems to reveal that their LLM implements a rather non-standard addition algorithm for integer arithmetic \cite{Anthropic2025c}.  Note also that the same report makes clear that when requested to explain its working, the LLM does not explain the actual algorithm seemingly used!

Charton has worked on explaining transformers designed for mathematics tasks \cite{Charton2022, LGCD_24}.  In a 2022 paper, they investigated the failure cases of transformers trained on matrix inversion and eigenvalue decomposition \cite{Charton2022}. He showed that incorrect model predictions still retain some mathematical properties of the solution: while models gave the right matrix 92\% of the time, in 99\% of examples the the matrix given had eigenvalues with less than 1\% error in the $L_1$ norm.  The conjecture arising from this is that mathematics transformers do not hallucinate absurd solutions. It also gives lessons on the dataset used: a smaller set with wider range of eigenvalues gives better generalisation.  Later, in his 2024 paper concerning a transformer model trained to compute the greatest common divisor, he also looked at explaining the performance \cite{LGCD_24}. The approach to explainability here was to engineer experiments to test hypothesis, rather than using any tools based on model parameters or input features.  This revealed some meaningful insights -- such as that  transformers learn to cluster input pairs with the same GCD and that early in training transformers learn to predict products of divisors of the base used to represent integers -- and so this method of explainability may offer a promising direction for other applications.

\section{Limitations and Future Possibilities of ML for CA}

It is evident that ML tools have played a role in optimising CA algorithms. The reported studies suggest that the practical computational efficacy of computationally intensive algorithms such as CAD and the computation of Gr\"{o}bner basis can be improved with the aid of ML tools, although care needs to be taken in the methodology used to achieve this. The variable ordering of CAD has become a well analysed case study in ML optimisation of CA, as described in Section \ref{CAD}.  The Gr\"{o}bner basis is perhaps the most commonly used CA tool for non-linear polynomials, but has only received study with ML more recently as described in Section \ref{gb}, perhaps due to its heuristic choices being more embedded within the algorithm and thus more difficult to study.  There is still scope for further work on both applications: for GB the state-of-the-art is limited in scope to certain ideal types, while for CAD there exists many heuristic decisions other that the variable ordering which have received little attention.

In Section \ref{SI} we saw that for symbolic integration metrics other than runtime were brought to the fore, with studies looking to optimise the form and interpretability of the output.  Although a variety of other areas of CA were found to have been studied by ML in the literature in Section \ref{OA}, it is clear that the applications that have been studied to date are only a small fraction of application areas within the field. The development of a shared methodology and toolkit would greatly aid developers to transfer knowledge between applications.

A common issue that arose is the inadequacy of benchmark dataset that can be used to make the proposed methodologies scalable and generalizable. It has long been accepted in CA that real-world applications data can look very different from the randomly generated datasets. Despite this, there exist few large-scale datasets and no community organisation controlling them.  For CAD, researchers took advantage of the SMT-LIB developed and controlled by the satisfiability checking community.  However, even this is not really of a scale for deep learning and it was found to need careful preparation before it can be used meaningfully.  In the absence of large real-world datasets, it is inevitable that synthetic data will be used for training, but in that case it should become a community standard of rigour that models are evaluated on real data.

There does not exist clear conclusions on which ML tools are the most effective for symbolic computation. As with most ML applications, there needs to be a tuning stage to the application at hand.  Recent work suggest the use of tree-based models \cite{SIA_24} and reinforcement learning \cite{SVO_24} as directions deserving future research.  

There exists a sharp choice between the use of ML to perform symbolic computation directly and its use to to optimise CA systems in a way that does not risk output correctness, as exemplified by \cite{LSS_20} and \cite{LCG_24} for Gr\"{o}bner basis computation.  The latter will likely generate new CA problems in algorithms to validate ML suggest results, and both approaches generate new CA problems in high quality synthetic dataset generation. It also worth noting that mathematics problems like these are particularly interesting for ML:  the difficulties in finding the right embedding and in identifying biases within datasets, make this a good testing ground for new ML innovations.  

Another key finding that emerged during this survey is that there are relatively few studies that use ML to entirely replace CA algorithms. Existing examples include work on symbolic integration \cite{DLS_19, KTYF22} and Gröbner basis computation \cite{LCG_24, HAT_25} discussed earlier, but nothing else was identified.

We finish by noting that another area deserving future focus is the use of explainable AI.  Most of the work surveyed gave no consideration to explainability but those that did suggested there is real value in doing so: both from new understanding of the algebra \cite{LSS_20}, from new understanding of how transformers learn \cite{LGCD_24}, and from better generalisability arising from simpler analysis suggested by XAI \cite{XAI_24}. By developing this and the other innovation described earlier, we may greatly increase the scope of applications for CA in real-world applications.

\begin{acks}
US is supported by a joint scholarship from Coventry University and Deakin University.  ME is partially supported by UKRI EPSRC project EP/T015748/1, \emph{Pushing Back the Doubly-Exponential Wall of Cylindrical Algebraic Decomposition} (DEWCAD).
\end{acks}

\bibliographystyle{ACM-Reference-Format}
\bibliography{Literature_review}

\appendix
\section{Appendix: Buchberger's Algorithm}
\label{app:BA}


Recall that we introduced Buchberger's Algorithm to compute a Gr\"{o}bner basis at the start of Section \ref{gb}.  

Suppose that we have a set of polynomials 
\[
F = \{ f_1, \dots, f_m \} \subset R[x_1, \dots, x_n]
\]
where $R[x_1, \dots, x_n]$ is a ring of polynomials. A Gr\"{o}bner basis 
\( G \) is computed as follows:

\begin{enumerate}
    \item \textbf{Initialise:} Set 
    \[
    G = F.
    \]

    \item \textbf{Compute $\mathbf{S}$-Polynomials:}
  
    For each pair \( (g_i, g_j) \in G \), compute
    \begin{equation}
    S(g_i, g_j) = 
    \frac{\operatorname{lcm}(\operatorname{LT}(g_i), \operatorname{LT}(g_j))}{\operatorname{LT}(g_i)} g_i 
    - \frac{\operatorname{lcm}(\operatorname{LT}(g_i), \operatorname{LT}(g_j))}{\operatorname{LT}(g_j)} g_j,
    \end{equation}
    where $\operatorname{lcm}$ stands for least common multiple and $\operatorname{LT}$ denotes the leading term.

    \item \textbf{Reduction of S-pairs:} If the $S$-pair \( S(g_i, g_j) \) is irreducible to zero modulo \( G \), it is added to \( G \) in its reduced form.

    \item \textbf{Repeat:} New elements keep being added by repeating the algorithm from Step 2 until no more $S$-polynomials produce non-zero reductions.

    \item \textbf{Output:} \( G \), a Gr\"{o}bner basis for $F$.
\end{enumerate}

\begin{figure}[h]
\centering
\includegraphics[width=\linewidth]{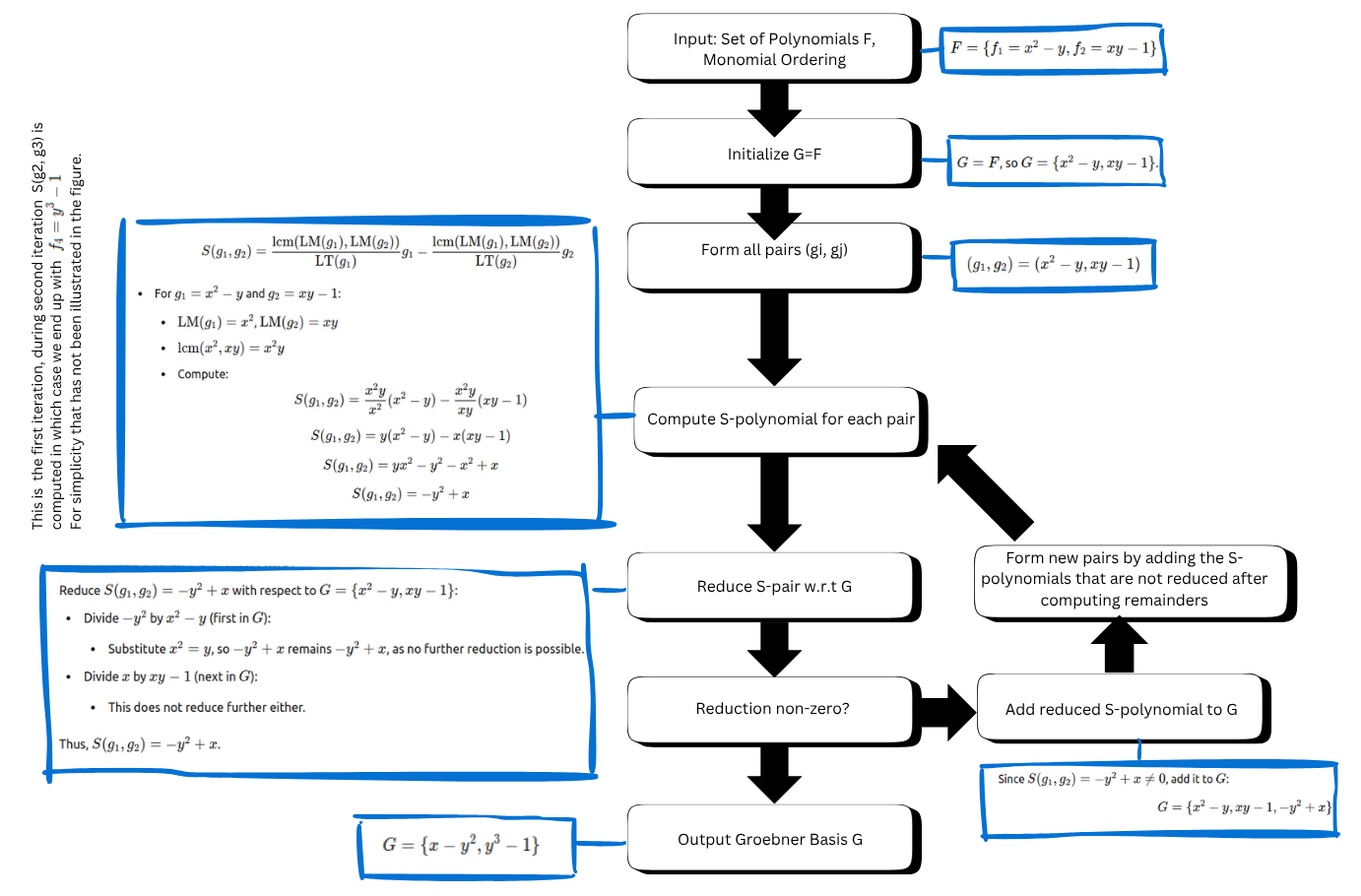}
\caption{\label{fig:flow}A flowchart showing Buchberger's Algorithm on an example.}
\Description{Buchberger's algorithm steps}
\end{figure}

Buchberger's algorithm is visualised in Figure~\ref{fig:flow} for an example set of polynomials.

\newpage

\section{Appendix: Summary Tables}
\label{app:ST}
\begin{table*}
  \caption{Summary of Studies on CAD and ML}
  \label{tab:CAD}
  \small 
  \begin{tabular}{p{0.14\textwidth} p{0.18\textwidth} p{0.18\textwidth} p{0.12\textwidth} p{0.08\textwidth} p{0.18\textwidth}}
    \toprule
    \textbf{Study} & \textbf{Focus Area} & \textbf{Key Findings} & \textbf{ML Method} & \textbf{XAI} & \textbf{Drawback} \\
    \midrule

    Huang et al. (2014) \cite{AMLP_14} & ML to choose from three human-designed heuristics for variable ordering. & ML choices better than any one heuristic. & SVM & No & Dataset limited to three variables; later found biased. \\

    Huang et al. (2016) \cite{UMLT_16} & ML to check if preconditioning with GB benefits solving non-linear systems with CAD. & On 1200 synthetic problems (2–4 vars), ML predicts when GB is beneficial. & SVM & No & Random dataset may not represent real-world data. \\

    England and Florescu (2019) \cite{CML_19} & Replaced heuristics in \cite{AMLP_14} with ML ordering directly. & Removes human entirely from ordering choice. & KNN, DT, MLP, SVM & No & ML ignores ranking of incorrect orderings. \\

    Florescu et al. (2019) \cite{AGNA_19} & Generated new features for CAD variable ordering. & outperformed human heuristics and benefitted from added features. & KNN, DT, MLP, SVM & No & Biased dataset. \\

    Florescu et al. (2020a) \cite{FE20-pipeline} & Open-source pipeline for CAD ordering. & Random datasets with stats similar to real-world inputs. & KNN, DT, MLP, SVM & No & Dataset realism not evaluated. \\

    Florescu et al. (2020b) \cite{ICV_20} & Cross-validation based on CAD runtime. & Sensitive to non-optimal orderings, improving reliability. & KNN, DT, MLP, SVM & No & Still framed as classification in training. \\

    Chen et al. (2020) \cite{CZC20} & Tackled fixed-variable limitation. & Iterative greedy selection + ML for final choice. & Deep Neural Network & No & --- \\

    Brown and Daves (2020) \cite{BD20} & Studied polynomial processing order in CAD. & Used repeated binary classification to form ranking. & Deep Neural Network & No & Limited by random dataset shape. \\

    del R{\'\i}o et al. (2023) \cite{DAM_23} & Uncovered bias in NLSAT dataset. & Data augmentation and balancing improved ML performance by 63\%. & KNN, DT, MLP, SVM & No & Data augmentation helps, but not optimal yet. \\

    del R{\'\i}o et al. (2024) \cite{LDP_24} & Reframed ordering as regression problem. & KNN, SVM perform better under regression; RF, MLP, XGB worse. & KNN, DT, MLP, SVM, XGB & No & Still supervised, data dependent. \\

    Jia et al. (2024) \cite{SVO_24} & Applied RL and GNN for ordering. & RL allows adaptive, data-driven selection. & RL, GNN & No & Advantage source unclear (RL vs GNN); performance drops on real. \\

    John et al. (2024) \cite{EAM_24} & Tested generalizability of extracted features. & Validated across original, balanced, augmented, shuffled datasets. & LR, KNN, DT, XGB, NN, GNN & Yes & Still supervised, sensitive to unseen data. \\

    Chen et al. (2024) \cite{DSVO_24} & Created large-scale dataset for CAD ordering. & First labelled dataset large enough for deep learning (20K + augmentation). & RF, MLP, Transformer & No & Data generated randomly; unclear representativeness. \\

    Pickering et al. (2024) \cite{XAI_24} & First XAI use (SHAP) in CAD ML. & Identified novel important features; XAI-informed heuristic performed strongly. & --- & Post-hoc (SHAP) & Based on \cite{AGNA_19} features; unclear if ML can generate better. \\

    Florescu and England (2024) \cite{CNNf_24} & Modelled Brown’s heuristic as constrained NN. & Trained on random, tested on real; strong performance. & Neural Network & Ante-hoc (small NN) & --- \\
    
    Jing et al. (2026) \cite{BDB_26} & 
    Pre-training and fine-tuning framework using six proxy algebraic tasks to learn deep polynomial structures. & 
    2$\times$ speedup on DQ-3; near-optimal (1.07$\times$ optimal) performance on real-world SMT benchmarks. & 
    Transformer & 
    No. & 
    low accuracy on deep symbolic operations like resultants \\

    \bottomrule
  \end{tabular}
\end{table*}

\begin{table*}
  \caption{Summary of Studies on Gr\"{o}bner Basis and ML}
  \label{tab:GB}
  \small
  \begin{tabular}{p{0.13\textwidth} p{0.15\textwidth} p{0.25\textwidth} p{0.13\textwidth} p{0.20\textwidth}}
    \toprule
    \textbf{Study} & \textbf{Focus Area} & \textbf{Key Findings} & \textbf{ML Method} & \textbf{Drawback} \\
    \midrule

    FMN (2020) \cite{FMN_20} & Approximate complex polynomial systems. & ML approximates systems to allow simpler Gr\"{o}bner basis computation. & Evolution strategies & Potential loss of structure; not suitable for non-permutation-invariant systems. \\

    Peifer et al. (2020) \cite{LSS_20} & Optimising Buchberger’s algorithm. & RL agent selects S-pairs for binomial systems, reducing computation. & Reinforcement Learning & Limited to binomials; trained and tested on random data. \\

    Mojsilovic \& Peifer (2021) \cite{LPM_22} & Predicting polynomial additions. & Simple features perform better than complex algebraic ones. & LR, RNN & Indirect improvement only; focuses on prediction, not algorithmic enhancement. \\

    Xu et al. (2024) \cite{OSI_19} & Variable ordering for Gr\"{o}bner basis. & Model selects variable order using coefficient sets. & Deep Learning & Overhead from ANN; applies only to permutation-invariant cases. \\

    Kera et al. (2024) \cite{LCG_24} & Gr\"{o}bner basis produced directly by transformers. & Hybrid embedding scheme for reverse Gröbner basis generation. & Transformers & Needs large datasets; fails on out-of-range coefficients; only approximate, no easy validation. \\

    Malhou et al. (2026) \cite{HAT_25} & Gr\"{o}bner basis computation using hierarchical attention transformers. & Tree-structured inductive bias matching polynomial hierarchy, combined with curriculum learning. & Hierarchical Attention Transformers & Still approximate with no easy validation; relies on curriculum learning to scale; performance beyond 13 variables unverified. \\

    \bottomrule
  \end{tabular}
\end{table*}

\begin{table*}
  \caption{Summary of Studies on Symbolic Integration and ML}
  \label{tab:symbolic_integration}
  \small
  \begin{tabular}{p{0.13\textwidth} p{0.15\textwidth} p{0.25\textwidth} p{0.13\textwidth} p{0.20\textwidth}}
    \toprule
    \textbf{Study} & \textbf{Focus Area} & \textbf{Key Findings} & \textbf{ML/RL Integration} & \textbf{Drawback} \\
    \midrule

    Lample and Charton (2020) \cite{DLS_19} & 
    Neural models for symbolic integration and differential equations. & 
    Transformers solve problems that CASs cannot. & 
    Seq2seq transformers & 
    Limited evidence of generalizability; dataset quality concerns. \\

    Albojwaid et al. (2023) \cite{SCD_23} & 
    Replication of \cite{DLS_19} with modified data generation. & 
    Larger training data, same methodology. & 
    Seq2seq transformers & 
    Little novelty compared to prior work. \\

    Barket et al. (2023) \cite{GEI_23} & 
    Dataset generation with elementary functions. & 
    Reverse-engineers Risch algorithm to address bias in \cite{DLS_19}. & 
    Data generation for ML & 
    Limited control over structure and length; restricted expression size. \\

    Barket et al. (2024) \cite{TLG_24} & 
    Dataset generation for elementary and non-elementary functions. & 
    Reverse-engineers Risch-Normal algorithm; adds variability and control. & 
    Data generation for ML & 
    Limited benchmarking of ML benefit; not extended to special functions. \\

    Barket et al. (2024) \cite{SIA_24} & 
    Predicting optimal Maple sub-algorithms via ML. & 
    Tree-LSTMs outperform standard LSTMs. & 
    LSTM and Tree-LSTM & 
    Does not generalise well enough to beat Maple on independent data. \\

    Barket et al. (2024) \cite{TPA_24} & 
    Predicting applicability of Maple methods via ML. & 
    Improves existing guards, adds new ones; XAI reveals missing simple guard rule. & 
    Seq2seq transformers; LIGs & 
    Most LIG outputs not interpretable; runtime impact of guards not evaluated. \\

    Sharma et al. (2024) \cite{sird_23} & 
    Selecting integration rules for functions. & 
    Large dataset enables interpretable integration paths. & 
    Step-level DNNs & 
    High computation required for multi-step learning. \\

    NNS (2022), MMP (2021) \cite{NNS_22}, \cite{MMP_21} & 
    University-level mathematics problems. & 
    Frontier models can perform symbolic integration. & 
    Yes & 
    Not benchmarked for CAS use; outputs not validated. \\

    Kubota et al. (2022) \cite{KTYF22} & 
    Critique of prefix notation use. & 
    Modified \cite{DLS_19} data; LSTMs outperform transformers via better token handling. & 
    LSTMs & 
    Not benchmarked for CAS use; outputs not validated. \\
    
    \bottomrule
  \end{tabular}
\end{table*}

\begin{table*}
  \caption{Summary of Studies on ML and Other Computer Algebra Tasks}
  \label{tab:combined-symbolic}
  \small
  \begin{tabular}{p{0.18\textwidth} p{0.28\textwidth} p{0.20\textwidth} p{0.20\textwidth}}
    \toprule
    \textbf{Study} & \textbf{Description} & \textbf{ML Role} & \textbf{Limitation} \\
    \midrule

    \multicolumn{4}{l}{\textbf{1. Using ML to guide symbolic algorithms}} \\[0.5em]

    Berczi et al. (2023) \cite{ARS_23} & Resolution of singular points as a game with RL agent. & Reinforcement Learning & Reward may be better tailored to problem. \\
    Hashemi et al. (2024) \cite{MLPS_24} & ML chooses coordinate changes for putting ideals into quasi-stable position. & Feature engineered supervised learning & Tested only on random ideals; generalizability unclear. \\

    \midrule
    \multicolumn{4}{l}{\textbf{2. ML algorithm selection for resultants}} \\[0.5em]

    Simpson et al. (2016) \cite{AASC_16} & Predicts best resultant algorithm in Maple/Mathematica for given instance. & Feature engineered supervised learning & Trained platform-specific; cross-platform performance uncertain. \\

    \midrule
    \multicolumn{4}{l}{\textbf{3. ML to improve CA for physics}} \\[0.5em]

    Alnuqaydan et al. (2023) \cite{SYMBA_23} & Transformer replaces symbolic squared amplitude step. & Seq-to-seq Transformers & Downstream effect of errors unclear. \\
    Kreuer et al. (2013; 2015) \cite{KPVv13, KUV15} & Horner form optimization for CA system FORM. & Heuristic Search + MCTS & Tailored to physics; not general-purpose. \\
    Ruijl et al. (2014) \cite{CSA_14} & Horner form optimization for CA system FORM. & Simulated Annealing + MCTS & Specific to FORM; not generalised CA use. \\

    \midrule
    \multicolumn{4}{l}{\textbf{4. Linear polynomial equation solving}} \\[0.5em]

    Dabelow \& Ueda (2024) \cite{SER_25} & Symbolic stack solver heuristic transformation choices. & Reinforcement Learning & Does not use CA systems; scalability unclear. \\
    Vaswani et al. (2024) \cite{VO24} & Transformer trained directly to solve linear systems. & Deep Learning Transformer & Black-box; lacks step-by-step explainability. \\

    \midrule
    \multicolumn{4}{l}{\textbf{5. Learning from the MATH-AI community}} \\[0.5em]

    Trinh et al. (2024) \cite{TWLHL24} & AlphaGeometry: solving Math Olympiad problems. & Hybrid RL + Deduction & Specific to geometry tasks. \\
    Wang et al. (2021) \cite{WLB21} & Recognising formula equivalence. & Neural networks; tree embeddings & Dependent on quality of synthetic data. \\
    Wankerl et al. (2021) \cite{LMR_21} & Recognising equivalence, derivative, or variable substitution in polynomials. & Neural networks; tree embedding & Focus on relations and computation. \\

    \midrule
    \multicolumn{4}{l}{\textbf{6. Explainable AI}} \\[0.5em]

    Charton (2022) \cite{Charton2022} & Analyses transformer errors in linear algebra tasks. & Interpretability via Error Properties & No input-level explainability. \\
    Charton (2024) \cite{LGCD_24} & Explains GCD prediction via structured experiments. & Experimental Explainability & Model internals still black-box. \\
    Anthropic (2025) \cite{Anthropic2025c} & Reveals inner workings of LLM for integer arithmetic. & Mechanistic Interpretability & Does not give full interpretations. \\

    \bottomrule
  \end{tabular}
\end{table*}

\end{document}